\documentclass[master]{ws-rv9x6} 
\usepackage{ws-rv-har}    
\usepackage{ws-rv-thm}    
\usepackage{subfigure}    
\usepackage{ws-index}     
\usepackage{enumitem}
\usepackage[percent]{overpic}
\usepackage{xcolor}
\makeindex
\newindex{aindx}{adx}{and}{Author Index}       
\renewindex{default}{idx}{ind}{Subject Index}  

\begin{document}		
\newcommand{\ltsima}{$\; \buildrel < \over \sim \;$}
\newcommand{\lsim}{\lower.5ex\hbox{\ltsima}}
\newcommand{\gtsima}{$\; \buildrel > \over \sim \;$}
\newcommand{\gsim}{\lower.5ex\hbox{\gtsima}}
\newcommand{\bra}{\langle}
\newcommand{\ket}{\rangle}
\newcommand{\lprime}{\ell^\prime}
\newcommand{\lpp}{\ell^{\prime\prime}}
\newcommand{\mprime}{m^\prime}
\newcommand{\mpp}{m^{\prime\prime}}
\newcommand{\ci}{\mathrm{i}}
\newcommand{\dd}{\mathrm{d}}
\newcommand{\veck}{\mathbf{k}}
\newcommand{\vecx}{\mathbf{x}}
\newcommand{\vecr}{\mathbf{r}}
\newcommand{\vecv}{\mathbf{\upsilon}}
\newcommand{\vecw}{\mathbf{\omega}}
\newcommand{\vecj}{\mathbf{j}}
\newcommand{\vecq}{\mathbf{q}}
\newcommand{\vecl}{\mathbf{l}}
\newcommand{\vecn}{\mathbf{n}}
\newcommand{\lm}{\ell m}
\newcommand{\that}{\hat{\theta}}
\newcommand{\thatp}{\that^\prime}
\newcommand{\chip}{\chi^\prime}
\newcommand{\hs}{\hspace{1mm}}
\newcommand{\nar}{New Astronomy Reviews}
\def\gsim{~\rlap{$>$}{\lower 1.0ex\hbox{$\sim$}}}
\def\lsim{~\rlap{$<$}{\lower 1.0ex\hbox{$\sim$}}}
\def\Msun {\,\mathrm{M}_\odot}
\def\Jcrit {J_\mathrm{crit}}
\newcommand{\rsun}{R_{\odot}}
\newcommand{\mbh}{M_{\rm BH}}
\newcommand{\Msunyr}{M_\odot~{\rm yr}^{-1}}
\newcommand{\mdot}{\dot{M}_*}
\newcommand{\ledd}{L_{\rm Edd}}
\newcommand{\cmc}{{\rm cm}^{-3}}
\def\gsim{~\rlap{$>$}{\lower 1.0ex\hbox{$\sim$}}}
\def\lsim{~\rlap{$<$}{\lower 1.0ex\hbox{$\sim$}}}
\def\Msun {\,\mathrm{M}_\odot}
\def\Jcrit {J_\mathrm{crit}}

\def\simgreat{\lower2pt\hbox{$\buildrel {\scriptstyle >}
   \over {\scriptstyle\sim}$}}
\def\simless{\lower2pt\hbox{$\buildrel {\scriptstyle <}
   \over {\scriptstyle\sim}$}}
\def\msobh{M_\bullet^{\rm sBH}}
\def\zodot{\,{\rm Z}_\odot}
\newcommand{\lambdabar}{\mbox{\makebox[-0.5ex][l]{$\lambda$} \raisebox{0.7ex}[0pt][0pt]{--}}}

\def\na{NewA}%
\def\aj{AJ}%
\def\araa{ARA\&A}%
\def\apj{ApJ}%
\def\apjl{ApJ}%
\def\jcap{JCAP}

\def\pasa{PASA}

\def\apjs{ApJS}%
\def\ao{Appl.~Opt.}%
\def\apss{Ap\&SS}%
\def\aap{A\&A}%
\def\aapr{A\&A~Rev.}%
\def\aaps{A\&AS}%
\def\azh{AZh}%
\def\baas{BAAS}%
\def\jrasc{JRASC}%
\def\memras{MmRAS}%
\def\mnras{MNRAS}%
\def\pra{Phys.~Rev.~A}%
\def\prb{Phys.~Rev.~B}%
\def\prc{Phys.~Rev.~C}%
\def\prd{Phys.~Rev.~D}%
\def\pre{Phys.~Rev.~E}%
\def\prl{Phys.~Rev.~Lett.}%
\def\pasp{PASP}%
\def\pasj{PASJ}%
\def\qjras{QJRAS}%
\def\skytel{S\&T}%
\def\solphys{Sol.~Phys.}%

\def\sovast{Soviet~Ast.}%
\def\ssr{Space~Sci.~Rev.}%
\def\zap{ZAp}%
\def\nat{Nature}%
\def\iaucirc{IAU~Circ.}%
\def\aplett{Astrophys.~Lett.}%
\def\apspr{Astrophys.~Space~Phys.~Res.}%
\def\bain{Bull.~Astron.~Inst.~Netherlands}%
\def\fcp{Fund.~Cosmic~Phys.}%
\def\gca{Geochim.~Cosmochim.~Acta}%
\def\grl{Geophys.~Res.~Lett.}%
\def\jcp{J.~Chem.~Phys.}%
\def\jgr{J.~Geophys.~Res.}%
\def\jqsrt{J.~Quant.~Spec.~Radiat.~Transf.}%
\def\memsai{Mem.~Soc.~Astron.~Italiana}%
\def\nphysa{Nucl.~Phys.~A}%

\def\physrep{Phys.~Rep.}%
\def\physscr{Phys.~Scr}%
\def\planss{Planet.~Space~Sci.}%
\def\procspie{Proc.~SPIE}%

\newcommand{\rmp}{Rev. Mod. Phys.}
\newcommand{\ijmpd}{Int. J. Mod. Phys. D}
\newcommand{\sovjetp}{Soviet J. Exp. Theor. Phys.}
\newcommand{\jkas}{J. Korean. Ast. Soc.}
\newcommand{\PPVI}{Protostars and Planets VI}
\newcommand{\njp}{New J. Phys.}
\newcommand{\rap}{Res. Astro. Astrophys.}

\setcounter{chapter}{2}
\chapter[Thermodynamics and Chemistry of the Early Universe]
{Thermodynamics and Chemistry of the Early Universe$^1$}
\footnotetext{$^1$ Preprint~of~a~review volume chapter to be published in Latif, M., \& Schleicher, D.R.G., ``Thermodynamics and Chemistry of the Early Universe'', Formation of the First Black Holes, 2018 \textcopyright Copyright World Scientific Publishing Company, https://www.worldscientific.com/worldscibooks/10.1142/10652.}
\label{chapter_chemistry}

\author[Stefano~Bovino \& Daniele~Galli]{Stefano Bovino$^1$ and Daniele Galli$^2$}
\address{$^1$Astronomy Department, \\Universidad de Concepci\'on, \\Barrio Universitario, \\ Concepci\'on, Chile,\\
stefanobovino@udec.cl}
\address{$^2$INAF -- Osservatorio Astrofisico di Arcetri, \\ Largo E. Fermi 5, I-50125 Firenze, Italy\\
galli@arcetri.astro.it}

\begin{abstract}
The interplay between chemistry and thermodynamics determines the
final outcome of the process of gravitational collapse and sets the
conditions for the formation of the first cosmological objects,
including primordial supermassive black holes. In this chapter we will
review the main chemical reactions and the most important heating/cooling processes taking place in a gas of primordial composition, including the
effects of local and cosmological radiation backgrounds.
\end{abstract}

\body


\section{Introduction}\label{intro}

The primordial environment out of which the first objects were formed is characterised by well established chemical initial conditions set by the recombination era, the absence of heavy elements (metals/dust), no turbulence before the formation of the first structures and no or very weak magnetic fields. For these reasons the early Universe can be considered a simplified model of the current interstellar medium (ISM). Molecular hydrogen is the main cooling agent in this scenario and its presence (or absence) can disentangle different dynamical paths. In this chapter we will present the most relevant processes affecting the chemistry of H$_2$ with particular focus on the main ions (H$^-$ and H$_2^+$) which determine its abundance. We will introduce the basic cooling/heating contributions which impact the thermodynamics of the gas and discuss high-energy processes (X-ray and cosmic-ray) that under certain conditions can influence the chemistry of the early Universe. Particular emphasis will be given to the radiation-induced processes which affect the dynamics of a primordial halo and its outcome. 

\subsection{Gas kinetics}

The motion and the behaviour of an ideal gas (fluid) is regulated by the momentum, the mass and the energy conservation laws.  These equations can be written both in Eulerian as well as in Lagrangian form (integral equations) where in the latter case a gas element is followed along its path.  In this section we will focus on the energy equation as we are interested on the thermal and chemical evolution of the gas.

The energy equation in Lagrangian notation reads as 
\begin{equation}
\label{eq:hydro}
	\rho\frac{D{\cal E}}{Dt} = -P\nabla\cdot {\bf u} 
	+  \Gamma (\rho, T_\mathrm{}) - \Lambda (\rho, T_\mathrm{}),
\end{equation}
where  ${\cal E}$ is the specific internal energy (in erg g$^{-1}$), $P= (\gamma -1)\rho {\cal E}$ is the gas pressure (with the adiabatic index $\gamma=5/3$ for  a monoatomic gas and $\gamma=7/5$ for a fully molecular gas),  ${\bf u}$ is the gas velocity,  $\Gamma$ and $\Lambda$ are the volumetric heating and cooling rates, respectively\footnote{Notice that the symbol $\Gamma$ is also often employed to represent photorates.}. In Eq.~(\ref{eq:hydro}) the
first term on the right-hand side represents the rate of doing $P\,dV$ work in adiabatic compressions of the gas. The heating and cooling rates are functions of the number densities $n_i$ of the  chemical species $i$ (represented here by the mass density $\rho$) and the gas temperature $T_\mathrm{}$. Both terms are strongly coupled to the chemical evolution of the gas, i.e. its kinetics.

Reactions between different species (ions, neutral, or photons) occur through collisions and their velocity/efficiency are regulated by the reaction rate coefficient $k$ that in general is a function of the gas temperature\footnote{In some specific cases the rate coefficient can also depend on other parameters (e.g. the density).}. The kinetics of a given species $i$, i.e. the {\em time} variation of its abundance $d n_i/dt$, is expressed via a {\em rate equation} that specifies the formation and destruction channels for that species. For example, let us consider a simple chemical system composed by two reactions, the photoionisation and the recombination of atomic hydrogen
\begin{eqnarray}
	\mathrm{H} + \gamma &\xrightarrow{k_1} & \mathrm{H}^+ + \mathrm{e}^-,\\
	\mathrm{H}^+ + \mathrm{e}^- &\xrightarrow{k_2} & \mathrm{H} + \gamma,
\end{eqnarray}
where $k_1$ and $k_2$ are the rate coefficients. Note that for a photochemical reaction the rate is expressed in s$^{-1}$ while standard reactions have units of cm$^{3(m-1)}$ s$^{-1}$, with $m$ representing the number of reactants; so, for example, for 2-body reactions ($m=2$) the units are cm$^3$ s$^{-1}$ while for 3-body reactions $m=3$ and the rate is in cm$^6$ s$^{-1}$.  In addition, the photoionisation rate $k_1$ is a function of the {\em radiation
temperature} $T_{\rm rad}$ characterising the radiation field. The rate equation for the evolution of hydrogen in this example is 
\begin{equation}
	\frac{dn_\mathrm{H}}{dt} = k_2 (T_{\rm })
	n_\mathrm{H^+}n_\mathrm{e} - k_1(T_{\rm rad}) n_\mathrm{H},
\end{equation} 
showing that, given an initial $n_\mathrm{H}$ abundance, it will increase as long as the {\em reaction flux} (in cm$^{-3}$ s$^{-1}$) for the recombination is larger than the photoionisation flux. The example above can be generalised to a multi-species case which includes any type of reactions, 
\begin{equation}\label{eq:kinetics}
	\frac{dn_i}{dt} = \sum_{j\in F_i} \left(k_j\prod_{r\in R_j}
	n_r(j)\right) -\sum_{j\in D_i} \left(k_j\prod_{r\in R_j}
	n_r(j)\right),
\end{equation} 
where the first sum includes all the terms which form the $i$-th species (belonging to the set $F_i$), while the second part is the analogous for the reactions that destroy the $i$-th species (set $D_i$). The $j$-th reaction has a set of reactants ($R_j$), and depends on the number density  $n_r(j)$ of each reactant at time $t$.

The energy equation (\ref{eq:hydro}) is usually solved by a standard operator-splitting technique, where the first term on the right-hand side is integrated alongside the mass and momentum conservation equations, while the energy change due to cooling and heating processes is coupled to the chemistry  (Eq.~\ref{eq:kinetics}) and added to the hydrodynamics as a source term. The thermal and chemical evolution together represent a stiff system of first-order ordinary differential equations (ODEs) that requires the use of accurate and adaptive numerical solvers~\citep{Bovino2013}. Besides their stiffness, astrophysical networks are also sparse (i.e. most of the Jacobian elements are zero) and this adds further challenges to an already expensive numerical problem. In addition, the system of ODEs can easily diverge when moving from zero to solar metallicity systems where metals, dust, and hundreds of species have to be taken into account to properly describe the kinetics of the gas. A reduced network of CO can already include $\sim 40$ species and $\sim 400$ reactions \citep{Glover2010,Grassi2017}.

\begin{figure}
\centerline{\includegraphics[scale=0.7]{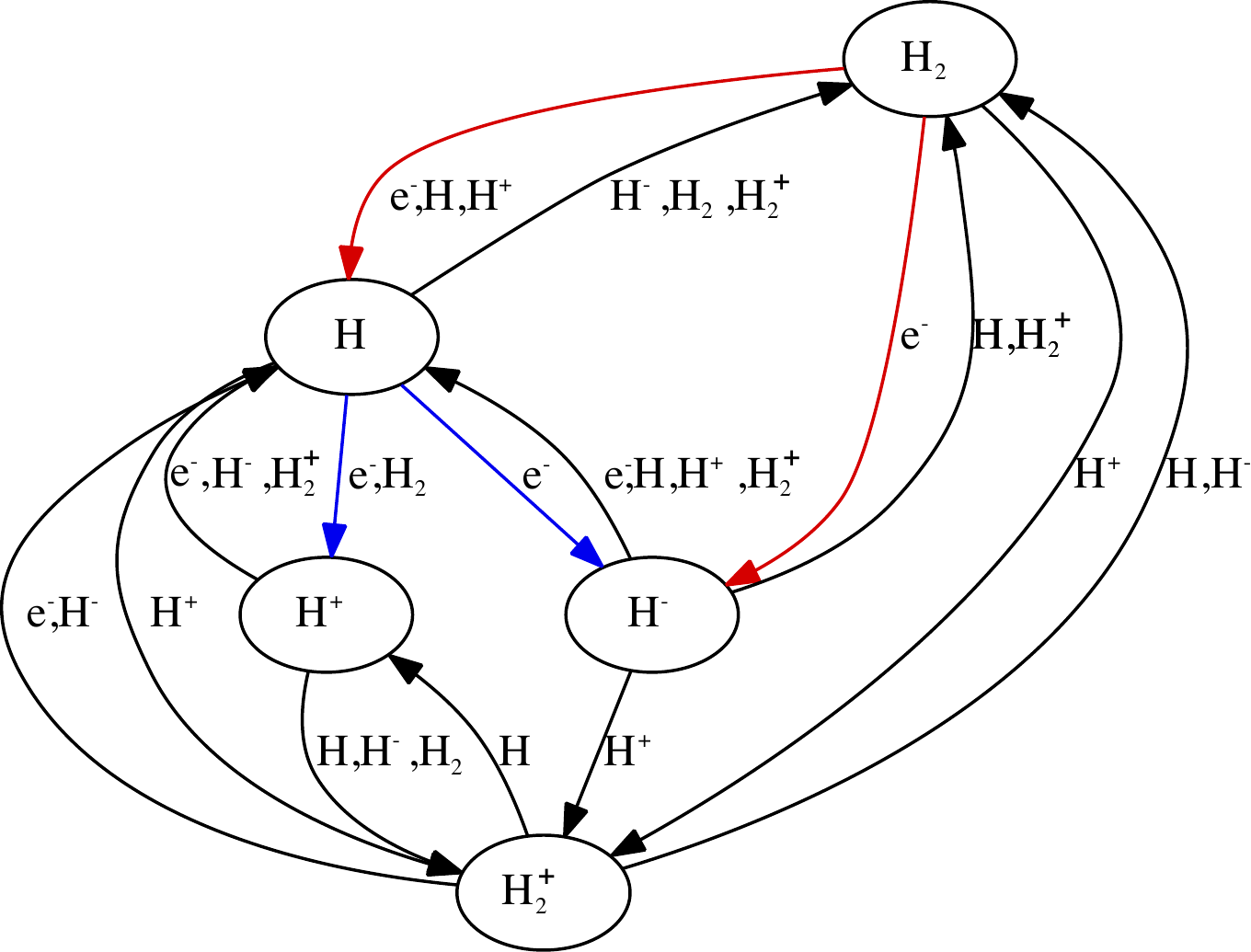}}
\caption{Graph of a simple H$_2$-oriented primordial network. Each arrow represents a reaction path where the reactants involved are sketched as hub or over the arrows. For example H$_2$ + e$^-$ produces H + H$^-$ as indicated by the red coloured arrows where H$_2$ is reported as a hub and electron as a reactant, or H + e$^-$ gives H$^+$ + H$^-$ as indicated by the blue coloured arrows. This graph has been obtained with the {\em docmake} tool released with the astrochemical package \textsc{krome} which makes use of the Graphviz software \mbox{(\protect\url{http://www.graphviz.org})}.}
\label{fig:network}
\end{figure}

Fortunately, primordial environments are relatively simple and can be considered as well established problems, with few uncertainties and a clear network of reactions.  An example is given in the graph reported in Fig.~\ref{fig:network}, showing a network of 9 species and roughly 20 reactions that can be employed for instance to accurately follow the formation of the first cosmological objects (see \citet{Latif2015} and references therein).

Following the chemical evolution of the primordial gas is very important because it allows to establish the connection between the microphysics (chemistry and cooling) and the macrophysics represented by the hydrodynamical equations (energy, mass, and momentum), and has a strong impact on relevant physical processes, as for instance the star-formation process. Chemistry is also needed when comparing theoretical work with observational data. However, chemical solvers can represent a bottleneck in modern numerical astrophysics. For this reason, fast and efficient astrochemical packages aimed at solving the chemical and thermal evolution of the gas have been developed in the last years, as for instance the \textsc{krome} package \citep{Grassi2014}.

\subsection{Ultraviolet radiation sources}
\label{sec:radiation}

Chemistry is also strongly connected to the radiation field, through photochemistry and photoheating, as discussed in this section. Photochemical processes in the early Universe are mainly induced by ultra-violet (UV) radiation produced by the first (Pop. III) and second (Pop.~II) generation of stars, and quasar-like ionising sources. In particular the Lyman-Werner (LW) background (at frequencies of 11.2--13.6~eV, corresponding to 912--1100~\AA) produced by Pop~III stars determines the onset of the soft UV background in the Universe and has a strong effect on the subsequent formation of stars \citep{Haiman1997,Omukai2001}.

\begin{figure}
\centerline{\includegraphics[scale=0.25]{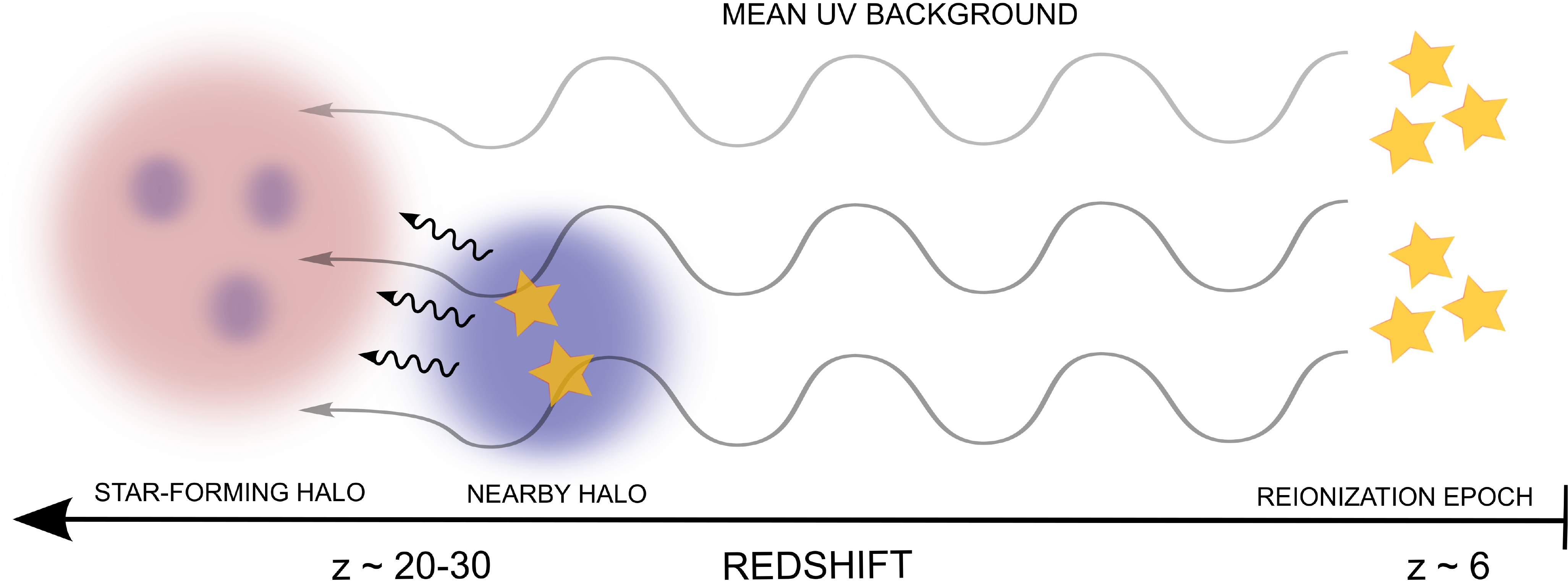}}
\caption{Pictorial representation of the different radiation sources which can irradiate a primordial star-forming halo (in pink) at high-redshift, namely the mean UV background from Pop. III stars emitted at the epoch of reionisation, and the strong UV radiation produced by Pop. III/Pop. II stars in a nearby halo. Redshifts and relative distances are indicative as well as the direction of the photons.}
\label{fig:radiation} 
\end{figure}

The total UV radiation that could be experienced by the gas in a dark-matter halo is the summation of the mean cosmic UV background produced just before the epoch of cosmic reionisation by massive stars \citep{Bromm03} and a local background coming from individual stellar sources hosted by nearby halos (see sketch in Fig.~\ref{fig:radiation}).  Population III stars in the host halo (galaxy) could also contribute to the amount of radiation together with the metal enrichment. As we are here interested in a metal-free environment the latter is not taken into consideration.

In the following we assume that most of the radiation above the Lyman limit (13.6 eV) is absorbed by HI, while LW photons can travel longer due to the low optical depth of the gas at those energies. The total LW background is then expressed as
\begin{equation}
	J_{\rm LW} = J_{\rm bg}^{\rm III} + J_{\rm loc}^{\rm III}
\end{equation}
where $J_{\rm loc}^{\rm III}$ is the local UV radiation coming from a nearby star-forming galaxy \citep{Visbal2014,Dijkstra08} produced by individual Pop. III stars, and $J_{\rm bg}^{\rm III}$ is the mean cosmological background, i.e. the minimum level of LW radiation that a halo is exposed to at any given redshift $z$.  The latter is usually defined as \citep{Greif2006,Omukai2001,Omukai2008}
\begin{equation}
	J_{\rm bg}^{\rm III} \approx \frac{1}{f_{\rm esc}} 
	\left(\frac{hc}{4 \pi m_\mathrm{H}}\right)  N_{\rm LW} \rho_*^{\rm III}(z) (1 + z)^3
\end{equation}
where $f_{\rm esc}$ is the escape fraction of LW photons from the halo, $N_{\rm LW}$ is the number of photons emitted in the LW bands per stellar baryon ($N_{\rm LW} \approx 2\times 10^4$, see \citep{Greif2006}), $h$ the Planck constant, $c$ the speed of light, $m_\mathrm{H}$ the proton mass, and $\rho_*^{\rm III}(z)$ the comoving density of Pop~III stars at a given redshift $z$. Similar expressions can be written for the radiation produced by
Pop~II stars\footnote{Throughout this book we express the radiation background in terms of the flux  $J_{21}$ at the Lyman limit (912~\AA, i.e. 13.6 eV): $J_{\rm LW} = J_{21} \times 
10^{-21}~\mbox{erg~s$^{-1}$~cm$^{-2}$~sr$^{-1}$~Hz$^{-1}$}$.}. Different approximations to the above spectra have been elaborated over the years. Quite often Pop~III and Pop~II spectra are approximated by uniform constant sources in terms of black-body radiation expressed as \mbox{$J_{\rm LW}(E) = J_{21}\times 10^{-21} B(E, T_{\star})/B(E_0,T_{\star})$}, where $E_0 =13.6$~eV, and $T_\star=10^5$~K and 10$^4$ K for Pop~III and Pop~II star, respectively.  Quasar-like sources are modelled as a simple power-law, $J_{\rm LW}(E) = J_{21}\times 10^{-21} (E/E_0)^{-1}$. Typical levels of LW background range from $J_{21}\approx 10^{-3}$ to $10^{-1}$, and depend on the stellar density at a given redshift \citep{Greif2006}, while local sources can have larger values $J_{21} > 10^3$.

Beside these simplified models, more realistic spectra have been proposed by different groups.  In \citet{Wolcott2017,Sugimura2014,Agarwal2015,Agarwal2016}, for example, more realistic spectra produced by Pop.~III and Pop.~II stars have been computed by employing stellar codes like STARBURST99. However these spectra should be taken with caution as they depend on many parameters/uncertainties, the most important being the initial mass function (IMF) of Pop. III stars which is difficult to constrain. The spatial and temporal dependency of the LW background has been considered by e.g. \citet{Visbal2014,Dijkstra08,Agarwal12}. Time-dependent average LW backgrounds have been proposed in simple parametrised form as a decreasing function of the redshift \citep{Visbal2014} and from more accurate calculations. 
The LW background is expected to increase with time \citep{Visbal2014} but it was established that neglecting time-dependency does not affect significantly the calculations of the final photorates, in particular as gravitational collapse will occur within a cosmologically short time span.

As we will see in the subsequent sections a UV radiation background in a primordial gas has the capability to photodissociate H$_2$ and H$_2^+$, and photodetach H$^-$. The shape of the assumed spectrum is strongly connected to the amplitude of the photorates for the radiatively-induced reactions, which are in turn relevant for the destruction/survival of molecules and ions in the primordial Universe. In the following section we will examine in detail the processes that affect the formation of the most important species in the early Universe, molecular hydrogen and its associated ions.

\section{Key species: H$_2$, H$^-$, and H$_2^+$ chemistry}

Basically, the chemistry of the early Universe reduces to three main species, H$_2$, H$^-$, and H$_2^+$, whose fates are strongly connected. Chemistry of deuterium is not discussed in this chapter but it is worth mentioning that it might be relevant for the formation of primordial low-mass stars through for example halo mergers \citep{Prieto2014,Bovino2014mer,Vasiliev2006} and the  cooling of post-shock gas \citep{Greif2006, Nakauchi2014}.

\subsection{{\rm H}$_2$ chemistry}\label{sec:h2chem}

H$_2$ is the simplest and the dominant molecule in the Universe and plays a crucial role in many astrophysical environments. It was the first species formed in the Universe and had great relevance in the formation of the first cosmological objects. Despite this, H$_2$ is characterised by peculiar chemical properties which affect its chemistry and the physics where it is involved. Being a homonuclear (symmetric) molecule, H$_2$ does not possesses a permanent dipole moment and the rovibrational transitions in its ground state occur through quadrupole radiation. As a first consequence, the formation of H$_2$ by direct radiative association of ground state H atoms is negligible because there are no allowed stabilising transitions. In present-day molecular clouds, H$_2$ is formed by surface reactions on dust grains that can efficiently absorb the energy which must be released in the process. The only catalysts available to form H$_2$ in zero-metallicity cloud are electrons and protons, rather minor constituents of the post-recombination Universe.

\paragraph{Main formation routes} 

The two routes to H$_2$ formation via gas-phase reactions in a cloud of primordial composition are
\begin{equation}
\label{eq:h2p}
{\rm H} + {\rm H}^+ \rightarrow {\rm H}_2^+ + h\nu,
\end{equation}
\begin{equation}
\label{eq:h2form_1}
{\rm H}_2^+ + {\rm H} \rightarrow {\rm H}_2 + {\rm H}^+
\end{equation}
(the H$_2^+$ channel) and 
\begin{equation}
\label{eq:hm}
{\rm H} + e^- \rightarrow {\rm H}^- + h\nu,
\end{equation}
\begin{equation}
\label{eq:h2form_2}
{\rm H}^- + {\rm H} \rightarrow {\rm H}_2 +e^-
\end{equation}
(the H$^-$ channel). Of these two routes, the latter is generally the most efficient. However, since H$^-$ has a smaller binding energy than H$^+_2$ (0.754~eV vs. 2.65~eV), it is more easily destroyed by any UV radiation released by Pop. III stars, minihalos and microquasars~\citep{Glover2007rad,Chuzhoy2007,Coppola2013}. The H$_2$ formed by these routes is destroyed mainly by photodissociation
\begin{equation}\label{eq:h2photo}
{\rm H}_2+ h\nu \rightarrow {\rm H} + {\rm H},
\end{equation}
and by collisional dissociation with H atoms
\begin{equation}\label{eq:colldiss}
{\rm H}_2+{\rm H} \rightarrow {\rm H} + {\rm H} + {\rm H}.
\end{equation}

\paragraph{Photodissociation} 

In spite of its low binding energy (4.48~eV), H$_2$ cannot be easily photodissociated, since direct excitation to the vibrational continuum of the ground electronic state $X^1\Sigma_g^+$ is forbidden by the dipole selection rules~\citep{Field1966}.  Direct excitation to the continua of higher electronic states is allowed, but requires absorption of a photon with energy larger than $14.7$~eV, which is more likely to photoionise H (threshold 13.6~eV) or H$_2$ (threshold at 15.5~eV). Thus, direct photodissociation is very inefficient \citep{Coppola2011}.  Photodissociation of H$_2$ mainly occurs by a two-step process (also called {\em Solomon process}): absorption of the Lyman ($X^1\Sigma_g^+\rightarrow B^1\Sigma_u^+$, threshold at 11.2~eV) or Werner ($X^1\Sigma_g^+\rightarrow C^1\Pi_u$, threshold at 12.3~eV) band UV photons, followed by radiative decay into the vibrational continuum of the ground electronic state \citep{Stecher1967,Glover2007rad}. On average, $\sim 15$\% of all radiative decays following absorption in the Lyman-Werner bands results in dissociation.  Since the Lyman-Werner bands arising from
the lowest rotational levels in the ground state comprise several hundred absorption lines between 912~\AA\ and 1100~\AA, the photodissociation rate of H$_2$ depends mainly on the intensity of the UV radiation in this wavelength range, quantified in terms of the $J_{21}$.

\paragraph{Collisional dissociation} 

While photodissociation is the main destruction process of H$_2$ in the presence of a significant UV background around $1000$~\AA, providing an efficient feedback mechanism to prevent further cooling and collapse \citep{Haiman1997,Haiman2000,Ciardi2000,machacek01,Mesinger2006}, collisional dissociation dominates at temperatures larger than 2000 K and in dense, shock-compressed regions \citep{Lepp1983,MacLow1986}. Typically, H$_2$ is completely dissociated by (non-magnetic) shocks with velocities above $\sim 20$~km~s$^{-1}$. The main contribution to collisional dissociation at low-densities is the \textit{dissociative tunneling}. This process involves quasi-bound states, i.e. excited states which are separated from the continuum by a rotational barrier, also called predissociative states. There is a high probability that molecules in these states can dissociate via tunneling instead to decay radiatively to a bound-state \citep{Martin1996}.

\paragraph{Three-body reactions} 

At densities above $\sim 10^{8}$~cm$^{-3}$, efficient H$_2$ formation is provided by three-body reactions \citep{Palla1983},
\begin{equation}
\label{eq:formation} 
{\rm H} + {\rm H} +{\rm H} \rightarrow {\rm H}_2 + {\rm H}, 
\end{equation}
\begin{equation}
\label{eq:3Bdiss} 
{\rm H} + {\rm H} +{\rm H}_2 \rightarrow {\rm H}_2 + {\rm H}_2, 
\end{equation}
the reverse processes of H$_2$ dissociation induced by collisions with H and H$_2$. For the impact of these reactions on the evolution of primordial clouds and the fragmentation process, see \citet{Turk2011a}, \citet{Bovino2014}.

\subsection{{\rm H}$_2^+$ chemistry}

In present-day molecular clouds, H$_2^+$ is formed by cosmic-ray ionisation of H$_2$ and rapidly transformed in H$_3^+$ by collision with other H$_2$ molecules.  Conversely, in the primordial gas H$_2^+$ is formed by radiative association (reaction \ref{eq:h2p}) in excited vibrational states, and is destroyed by the reverse reaction, photodissociation, at a rate strongly dependent on the distribution of excited levels \citep{Coppola2011}. In its ground vibrational state, H$_2^+$ is photodissociated by photons with $h\nu > 2.65$~eV, or $\lambda < 4680$~\AA, and therefore survives better than H$^-$ to intense UV radiation fields. In those circumstances, H$_2^+$ represents the main source of H$_2$ via the proton exchange reaction (4), as, e.g. in the post-recombination Universe at redshifts $z\gtrsim 300$ \citep{Saslaw1967,Galli1998}. However, when the gas is sufficiently ionised, H$_2^+$ is promptly removed
by dissociative recombination with electrons, a reaction enhanced by the presence of several low-energy resonances \citep{Takagi2002,Schneider2002}.

\subsection{{\rm H}$^-$ chemistry}

The hydrogen anion H$^-$ (hydride ion) is the main catalyst for the formation of H$_2$ in the primordial gas. This species, a proton with two bound electrons, has a binding energy of only $0.754$~eV and no bound excited states.  In the present-day Universe, it provides the continuum opacity of the Sun and most other late-type stars via bound-free and free-free transitions; similarly, in the post-recombination Universe, it represents the main source of opacity for the cosmic microwave background radiation, although at levels presently undetectable \citep{Black2006,Schleicher2008}.  In the primordial gas, H$^-$ is formed by radiative attachment (reaction~\ref{eq:hm}) and removed by associative detachment (reaction~\ref{eq:h2form_2}). Recently, reaction (\ref{eq:hm}) has been the object of intensive laboratory experiments and theoretical calculations that have largely reduced the uncertainties on the reaction rate of this crucial process \citep{Bruhns2010,Kreckel2010,Miller2011}.

\paragraph{Photodetachment} 

In the presence of a significant flux of photons with $h\nu >0.754$~eV ($\lambda < 16,500$~\AA), H$^-$ is efficiently destroyed by  {\em photodetachment} (reverse of reaction~\ref{eq:hm}).  The cross-section of this process has been calculated by different  groups with~\citep{Miyake2010} and without~\citep{Wishart1979} resonant contributions (for photon energies greater than $\sim 11$ eV).  The photodissociation rate is also strongly affected by the shape of the UV spectrum, as for instance a typical power-law radiation background produced by miniquasar can boost the rate by a factor of $\sim5$~\citep{Chuzhoy2007}. In general the net effect of the photodetachment process is to suppress the formation of H$_2$ reducing the efficiency of the H$^-$ channel by a factor of approximately $1 + k_{\rm ph} /(k_{\rm ad} n_\mathrm{H})$, where $k_{\rm ph}$ is the photodetachment rate (reverse of action~\ref{eq:hm}), 
$k_{\rm ad}$ the associative detachment rate (reaction~\ref{eq:h2form_2}),  and $n_\mathrm{H}$ the number density of atomic hydrogen (see e.g. \citet{Chuzhoy2007,Miyake2010}). Recently has been also suggested that trapped Ly-$\alpha$ cooling radiation emitted from collapsing gas (see sec. \ref{sec:lyman}) could contribute to the photodetachment of H$^-$. The rate for this process has been provided in \citet{Johnson2016} as a function of the cloud mass, the density, and the temperature of the gas.

\section{High-energy chemistry}

As mentioned in Sect. \ref{sec:h2chem}, the main cooling agent in the primordial gas, H$_2$, is efficiently photodissociated by photons with energies in the range of  the Lyman-Werner bands (11.2--13.6~eV). Since these photons are not strongly absorbed by ambient H atoms, it is natural to expect the formation of the very first massive stars to be accompanied by the generation of a strong UV background radiation field (Sect.~\ref{sec:radiation}), which is able in
principle to photodissociate any residual H$_2$ and suppress further star formation (see Sect. \ref{sec:h2chem} and reference therein). Ionisation of H$_2$ by UV photons, on the other hand, is relatively inefficient because ionising photons with energy above the threshold energy of 15.4~eV only exist inside HII regions.

In contrast, X-rays ($h\nu > 1$ keV) and, to a larger extent, cosmic rays ($E > 1$ MeV) can penetrate protogalactic clouds and ionise both H and H$_2$,  producing ions and excited molecules that can interact with the neutral atomic or molecular gas (see, e.g. \citealt{Tielens2005}). The ionisation of atomic hydrogen generates electrons that can catalyse further formation of H$_2$, compensating, or even overcoming the destruction of H$_2$ by the same process.
These reactions carry away a significant amount of the available energy in the form of chemical heating, and contribute to the heating of the gas as much as the ionisation electrons themselves \citep{Goldsmith1978,Glassgold2012}.

Possible sources of an X-ray background in the early Universe include high-redshift quasars, massive X-ray binaries, Bremsstrahlung and inverse Compton emission from Pop.~III supernova remnants \citep{Haiman2000,Glover2003,Hummel2015,Latif2015}. The latter are also the most likely sources of cosmic-ray particles,  although more exotic processes like the decay of supermassive particles are also possible \citep{Shchekinov2004,Ripamonti2007MNRAS,Jasche2007,Hummel2016}. In general, cosmic rays penetrate deeper into a cloud than X-rays, but their relative importance as ionising and heating agents depends on poorly constrained quantities like the intensity, distribution and spectral characteristics of the sources. Numerical models show that the effects of an X-ray background on the chemical and thermal evolution of protogalactic clouds could be significant at relatively low densities, while the high-density, star-forming gas remains largely unaffected \citep{Hummel2016}.

\section{Shielding effects}

When considering a cloud irradiated by a radiation background it is very important to take possible {\em shielding} effects into account. The photodissociation rates and the cooling functions vary with the gas column density and only high energy photons and cosmic rays can penetrate deep into the cloud (where column densities are high). In present-day molecular clouds photons can be shielded both by the gas as well as dust, the latter being the main contribution for energies
below the Lyman limit. 

In a primordial environment the most important process is the H$_2$ {\em self-shielding}. At high enough densities H$_2$ molecules gather together creating a thick layer of material which can absorb energetic photons decreasing the photodissociation rate. This is usually taken into account by a suppression factor $f_{\rm sh}$, so that the rate is expressed as \mbox{$k_{\rm ph}^{\rm thick}= f_{\rm sh} k_{\rm ph}^{\rm thin}$}, where $f_{\rm sh} = 1$ is the ``no-shielding" case (optically thin regime), while $f_{\rm sh} = 0$ is a complete shielding, i. e. radiation fully absorbed.

Evaluating the shielding is non-trivial as it usually requires the solution of the radiative transport equations and the computation of the gas column density. This depends on geometry, density, temperature, velocity gradients, and the presence of turbulence.  Accurate calculations have been performed over the years and simple analytical formulae have been provided \citep{Draine1996,WolcottGreen2011,Safranek2012,Hartwig2015} based on approximate definitions of the column density. The latter is frequently expressed based on local properties as $N_i = n_i \ell$, with $n_i$ being the number density of the considered species (in a primordial cloud usually H$_2$) and $\ell$ a characteristic length.
Its choice however depends strongly on the physical problem, and this approximation can fail to reproduce accurate radiative transfer calculations under particular conditions. Detailed comparisons between the different characteristic lengths have been provided by several authors \citep{WolcottGreen2011,Safranek2012,Hartwig2015}.

When calculating the self-shielding one also has to take into account turbulence and thermal Doppler broadening~\citep{Richings2014}. The latter is important for instance in the study of collapsing gas where the self-shielding of H$_2$ molecules occurs only if relative velocities $u_r$ are  smaller than the thermal velocity, $u_r \ll v_\mathrm{th}$. An accurate method to account for this effect has been developed by \citet{Hartwig2015}.

To conclude, changes in the H$_2$ self-shielding treatment can introduce up to an order of magnitude differences on the final photodissociation rate, affecting in equal measure the outcome of numerical simulations of first cosmological objects.

\section{Thermodynamics in primordial environments}

A fundamental ingredient in any cosmological scenario for the formation of the first stars is the ability of primordial clouds to radiate energy efficiently during the phase of gravitational collapse.  Clearly, the main cooling mechanisms at work in present-day clouds, namely CO and metal line emission or IR emission from dust grains are not available in a cloud of zero metallicity.

\subsection{Cooling processes}

In a H/He plasma in ionisation equilibrium, thermal kinetic energy can be removed from the gas by collisional excitation/ionisation, recombination, and radiation of free electrons scattering or free ions (free-free emission or Bremsstrahlung). The latter is in fact the primary source of cooling for $T_{\rm gas}\gtrsim 10^6$~K, when the gas is fully ionised, but its efficiency decreases with decreasing temperature. At $T_{\rm }\approx 10^5$~K, and $T_{\rm }\approx 5\times 10^4$~K, the cooling is dominated by collisional excitation of He$^+$ and H, respectively, followed by radiative decay (mostly in the Lyman $\alpha$ line of neutral hydrogen). Below $T_{\rm }\approx 10^4$~K, the plasma recombines very rapidly. The very few residual electrons are unable to excite the first excited level of H, with an energy 10.2~eV above the ground state, and the cooling rate drops to zero.  The cloud then relaxes to a stable thermal state with $T_{\rm }\approx 10^4$~K. While in a gas of non-zero metallicity, oxygen and nitrogen ions with the lowest energy levels just a few eV above the ground state can further cool the gas, the thermal evolution of a primordial cloud is controlled by the amount of H$_2$ and HD, and, to a lesser extent, H$_3^+$, formed by gas-phase reactions in a dust-free environment \citep{Saslaw1967,Matsuda1969}.

\subsubsection{{\rm H}$_2$ cooling}\label{sec:h2cool}

The cooling of a metal-free cloud is dominated by the spontaneous emission of rovibrational levels of H$_2$ collisionally excited mainly by H and He atoms or other H$_2$ molecules. In the optically thin regime, the emitted photons can freely escape the cloud without being re-absorbed or scattered, transferring thermal energy outward at a rate proportional to the H$_2$ abundance.  However, despite being the most abundant molecular species formed by gas-phase reactions in a zero-metal gas, H$_2$ is not an efficient coolant. Due to its lack of a permanent electric dipole moment, rovibrational levels of H$_2$ have small transition probabilities, and can be collisionally de-excited even at relatively low densities. In fact, above a {\em critical density} $n_{\rm crit}\approx 10^4$~cm$^{-3}$, the H$_2$ cooling rate reaches its local thermodynamic equilibrium (LTE) value, given by
\begin{equation}
\Lambda_{\rm LTE}=\sum_{u,l} n_u A_{ul} E_{ul},
\end{equation}
where $n_u$ is the population density of H$_2$ in upper energy level $u$, $A_{ul}$ is the transition probability for spontaneous decay to energy level $l$, and $E_{ul}=h\nu_{ul}$. Note that typically, $n_\mathrm{crit}\approx 10^4$ cm$^{-3}$ for the lower rotational states \citep{LeBourlot1999} but the vibrational states can have a higher $n_\mathrm{crit}$ of about \mbox{10$^8$ cm$^{-3}$}. However, high vibrational levels will have a negligible population already
at 8000 K (only the first 2-3 levels will be populated but the ground will still dominate). Useful formulae for the H$_2$ cooling in the low-density and LTE conditions can be found in \citet{Glover2008} and \citet{Coppola2012}, respectively.  The former reference also contains a thorough discussion of the H$_2$ ortho/para ratio in the typical conditions of primordial clouds, where it is generally different from the equilibrium value 3:1 expected at high densities and temperatures.

At densities above $n_{\rm thick}\approx 10^9$~cm$^{-3}$, H$_2$ becomes optically thick to its own radiation, and the emitted rovibrational photons can be efficiently scattered and re-absorbed, decreasing the overall cooling efficiency.  However, velocity gradients can Doppler-shift photons outside the optically thick cores of the rovibrational lines, allowing them to leave the cloud \citep{Yoshida2006,Stacy2013,Hartwig2015ApJ}.  In this case, one usually adopts the Sobolev approximation, assuming that each radiative transition probability $A_{ul}$ is reduced by a factor $\beta_{ul}$, the so-called ``escape probability'', that depends on the velocity profile of the cloud \citep{Yoshida2008,Clark2011b,Greif2011,Greif12}.

\subsubsection{{\rm HD} cooling}

In spite of a cosmological D/H ratio of $\sim 10^{-5}$, the role of HD as a cooling agent is not negligible.  First, the HD molecule has a small permanent electric dipole moment and radiative transition probabilities $\sim 100$ times larger than those of H$_2$. Therefore, the critical density for the onset of LTE in the case of HD cooling is larger than that for H$_2$, of the order of $n_{\rm crit}\approx 10^6$~cm$^{-3}$. Second, because of its larger mass, the energy difference of the lowest rotational transition of HD corresponds to a temperature of 128~K, rather than 510~K as in the case of H$_2$, and therefore can efficiently cool a gas to lower temperatures than H$_2$. Third, since HD is mostly formed by the reaction
\begin{equation}
{\rm D}^+ +{\rm H}_2 \rightarrow {\rm HD} + {\rm H}^+, 
\end{equation}
which is exothermic by 462~K, its abundance relative to H$_2$ is boosted at low temperature as \mbox{$[{\rm HD}/{\rm H}_2]\approx 2[{\rm D}/{\rm H}]\exp(462~{\rm K}/T_{\rm })$} (a process called ``deuterium fractionation''). Thus, HD is able to cool the gas to temperatures below 200~K, possibly all the way to that of the cosmic background, which sets a lower limit to radiative cooling \citep{Johnson2006,Glover2008}. Useful formulae for HD cooling for collisions with H are given by \citet{Lipovka2005} and \citet{Coppola2011} in the low-density and LTE limits, respectively.

In general, HD cooling plays a marginal role for the evolution of typical primordial halo masses ($M\approx 10^6~M_\odot$), where the temperature never falls below $\sim 200$~K \citep{Bromm2002,Yoshida2006}. However, HD cooling can affect significantly the collapse of low-mass halos ($M\lesssim 3\times 10^5~M_\odot$), leading to a less massive stellar population compared to the case of halos dominated by H$_2$ cooling  \citep{Ripamonti2007MNRAS,McGreer2008}.  In addition, HD cooling becomes important in all situations where the gas is initially significantly ionised, as in halos formed behind strong shocks \citep{Shapiro1987,Johnson2006,Greif2006, Nakauchi2014} as in merger events \citep{Prieto2014,Bovino2014} or in relic HII regions \citep{Mackey2003,Johnson2006}.

\begin{figure}
\centerline{\includegraphics[scale=0.4]{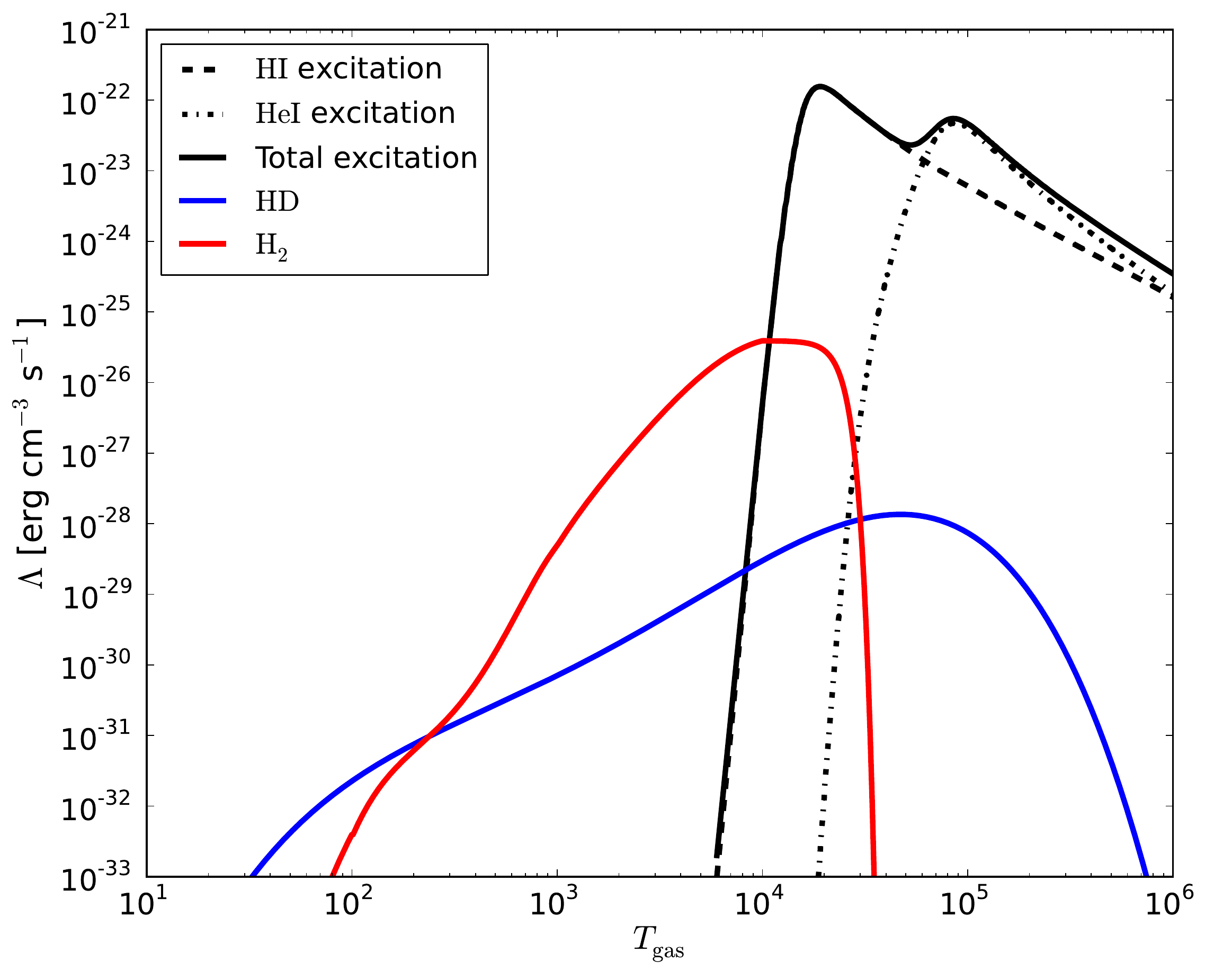}}
\caption{Cooling functions due to hydrogen (HI), helium (HeI) excitations, molecular hydrogen H$_2$, and HD as a function of the gas temperature for a given density \mbox{$n = 1$ cm$^{-3}$}. The H$_2$, and HD cooling functions are computed at fixed species number fractions, i.e. $x_\mathrm{H_2} = 10^{-4}$, and $x_\mathrm{HD} = 10^{-7}$. The atomic cooling is instead calculated assuming collisional ionisation equilibrium (see \citet{Katz1996}).}
\label{fig:cooling}
\end{figure}

A comparison between HD and H$_2$ cooling is reported in Fig.~\ref{fig:cooling} together with H and He atomic cooling.

\subsubsection{Ly-$\alpha$ cooling}\label{sec:lyman}

The energy levels of hydrogen atoms mainly depend on two quantum numbers, the electronic quantum number $n$, and the orbital quantum number $l$, with energy scaling as 13.6 eV/$n^2$. We restrict our discussion here to the first three electronic levels, termed $1s$ ($n=0,l=0$), $2s$ ($n=1,l=0$), and $2p$ ($n=1,l=1$). Spontaneous transitions between \mbox{$2s$--$1s$} are forbidden by selection rules, while transitions involving the ground state $n=1$ give rise to the Lyman series\footnote{When the final state is $n=3$, or $n=2$ we refer to the Paschen and Balmer series, not of interest in this chapter.}.

\begin{figure}
\centerline{\includegraphics[scale=0.8]{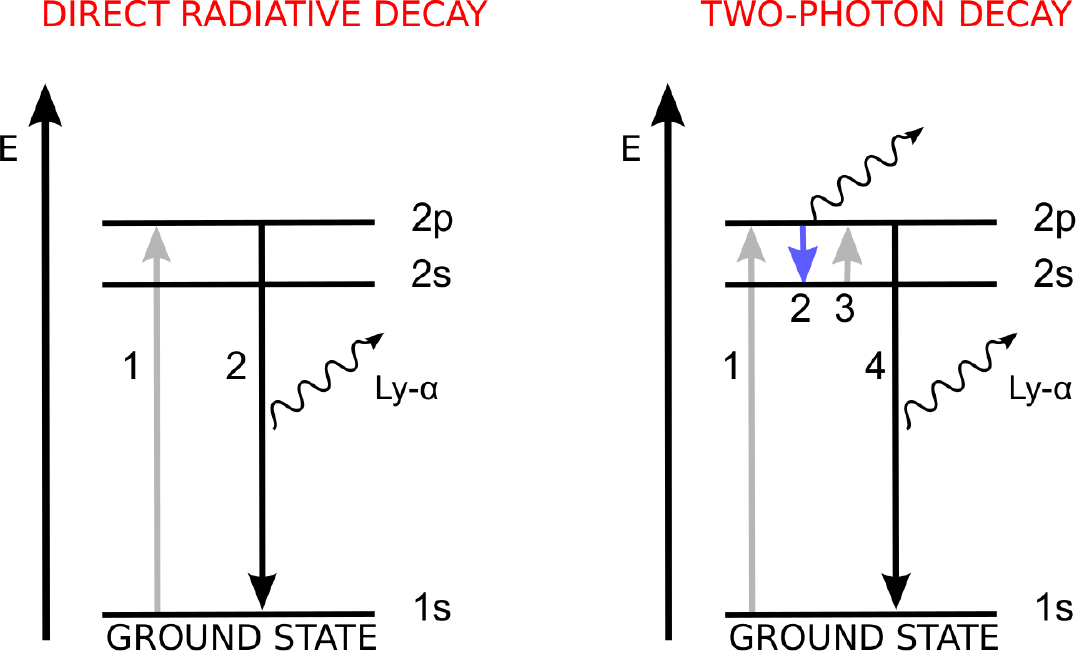}} 
\caption{Sketch of the hydrogen Ly-$\alpha$ cooling and the two-photon process. We only show the first three hydrogen levels. Numbers represent the sequence of events. The grey arrows indicate the excitation by collisions with electrons; the black arrow the photon emission that produces Ly-$\alpha$ photons; the blue arrow the decay from $2p$ to $2s$.}
\label{fig:lyman}
\end{figure}

The so called Ly-$\alpha$ cooling originates from the collisional excitation of hydrogen atoms from the ground electronic state to the $2p$ state, which in the subsequent emission process produces Ly-$\alpha$ photons at 10.2 eV (left sketch in Fig. \ref{fig:lyman}). In metal-free cloud this contribution is crucial to bring the gas down to 8000~K.  The cooling function is usually expressed as $\Lambda^{{\rm Ly}\alpha} =  n_{\rm e} \sum_i\sum_u n_i E_{ul}^i \langle\sigma_{ul}^i u_{\rm e}\rangle$,  with $n_{\rm e}$ the electron density, $n_u$ and $n_l$ the density of the upper and lower levels, $E_{ul}$ the energy difference between the two levels, $\sigma_{ul}$ the excitation cross-section, and $u_{\rm e}$ the electron velocity. Note that the cross-section is averaged over a Maxwellian distribution~\citep{Shapiro1987}.

\subsubsection{Two-photon decay} 

Another process which can contribute in part to the cooling of the gas producing Ly-$\alpha$ photons is the so called two-photon decay \citep{Spitzer1951,Omukai2001,Glover2007rad}. Even though the $2s$--$1s$ transition is forbidden there is a chance that hydrogen atoms (after being excited through collisions with electrons into the $2p$ state) decay to the $2s$ state and are again excited to the $2p$ followed by the decay to the ground state (see sketch in Fig. \ref{fig:lyman}, right). The sum of the two emitted photons is a net Ly-$\alpha$ transition at 10.2 eV. This process is in general $\sim$10$^8$  orders of magnitude slower than the Ly-$\alpha$ direct radiative decay and depends on different parameters as for instance the Ly-$\alpha$ optical depth of the gas. An analytical formula of this cooling is given for example in \citet{Omukai2001}.

\subsubsection{High-density cooling}

When densities increase, H$_2$ reaches LTE and its capability to cool the gas is highly reduced. Three-body reactions then trigger the cooling at density between $10^8$--$10^{10}$~cm$^{-3}$, efficiently contributing to the conversion of hydrogen atoms to H$_2$ molecules. When densities reach $\sim 10^{14}$~cm$^{-3}$ the gas becomes optically thick to H$_2$ and the H$_2$ cooling is completely inefficient.  In the high density regime between
$10^{14}$--$10^{17}$~cm$^{-3}$  other processes contribute to the thermodynamics of the gas, among these the collisional induced emission. In the following we discuss the main contributions in the density regime where H$_2$ cooling becomes inefficient, focusing in particular on the cooling produced by chemical reactions and by continuum processes.

\paragraph{Chemical cooling}

A chemical reaction proceeds in time along its reaction path going from reactants to products through different mechanisms. We can distinguish between two main classes of reactions: ({\em i}\/) \textit{exothermic}, and ({\em ii}\/) \textit{endothermic}. The former ones proceed fast and release into the medium a given amount of energy equal to the enthalpy of the reaction $\Delta H$. On the contrary, endothermic reactions need to absorb energy from the
medium to take place. The outcome of endothermic reactions is a net cooling effect, with cooling function proportional to the energy needed for the reaction to occurs ($\Delta H$), the abundances of the reactants ($i, j$), and the reaction rate $k(T_{\rm })$, which regulates the speed of the entire process~\footnote{For reactions involving a third body the expression should read as $n_i n_j n_k k(T_{\rm }) \Delta H$.}. The chemical heating 
$\Lambda_\mathrm{chem} \propto n_i n_j k(T_{\rm }) \Delta H$ is expressed in erg~cm$^{-3}$~s$^{-1}$.

In a primordial environment the main source of chemical cooling comes from the destruction of H$_2$ via three-body reactions (reaction~\ref{eq:3Bdiss}) and collisional dissociation (reaction~\ref{eq:colldiss}), and results in 4.48 eV 
carried away from the medium. These processes have an impact on the thermal evolution of a collapsing gas at density larger than 10$^8$~cm$^{-3}$.

\paragraph{Continuum processes} The continuum cooling is usually expressed as $\Lambda_\mathrm{cont} = 4\sigma T_{\rm }^4 \kappa_\mathrm{gas} \beta_\mathrm{esc}$, i.e. photon emission based on the Stefan-Boltzmann law (with $\sigma$ the Stefan-Boltzmann constant) but attenuated by the opacity of the gas $\kappa_\mathrm{gas}$ and the escape probability $\beta_\mathrm{esc}$ introduced in sec. \ref{sec:h2cool}.  The main processes which contribute to the opacity of the gas at high-densities are bound-free absorption of H, He, H$^-$, and H$_2^+$ ; free-free absorption of H$^-$ and H; collision-induced absorption of H$_2$--H$_2$ and H$_2$--He; Rayleigh scattering of H; and Thomson scattering of electrons  \citep{Lenzuni1991}.

The main contribution is provided by collisionally-induced absorption (emission) of H$_2$~\citep{Lenzuni1991,Ripamonti2004,Yoshida2008,Hirano2013}. At high densities the number of collisions is highly enhanced; in particular H$_2$ can inelastically collide with other H$_2$ molecules or helium atoms inducing the formation of a transient supramolecule with a non-zero dipole moment, with a high probability of emitting a photon. This is the last radiative cooling channel available during the collapse of a primordial cloud. Above 10$^{16}$ cm$^{-3}$ the gas becomes completely opaque to the continuum radiation and also  collisionally-induced emission becomes inefficient \citep{Yoshida2008}.

\subsection{Heating processes}

There are different ways to heat the gas in the ISM, i.e. to transfer energy into the gaseous medium: ({\em i}\/) via photons/electrons, ({\em ii}\/) via chemical reactions, ({\em iii}\/) via dust grains, and ({\em iv}\/) via dynamical processes. In this section we will focus on the microscopic processes relevant for a primordial environment, mainly the chemical heating and the photoheating. However, other processes at macroscales can strongly contribute to
the heating, e.g. dynamical compression (first term in Eq.~\ref{eq:hydro}) during gravitational collapse, shocks and turbulence, mechanical heating produced by stellar winds and supernova explosion, and viscous heating (e.g. ambipolar diffusion). For an overview on these dynamical processes we refer to \citet{Tielens2005}.

\subsubsection{Chemical Heating}
\label{sec:heatchem} 

Exothermic reactions, i.e. processes which proceed very fast, often without a barrier, and with an excess of energy, can contribute to the heating of the gas. In this class of reactions, collisions between atoms and molecules are usually converted into ({\em i}\/) translation energy of the newly formed molecule, and ({\em ii}\/) rotational/vibrational excitations. In the former case the heating is directly released into the gas, in the latter the gas is heated up via subsequent de-excitations.

In a primordial gas chemical heating is mainly due to processes which form molecular hydrogen, in particular the H$^-$ channel, the H$_2^+$ channel, and three-body reactions. The heat deposited per formed molecular hydrogen is defined similarly to the chemical cooling as \mbox{$\Gamma_\mathrm{chem} = \epsilon n_i n_j k(T_{\rm }) \Delta H$}, where $\Delta H=4.48$~eV, 3.53~eV, and 1.83~eV respectively for the reactions cited above. At variance with chemical cooling, chemical heating is weighted by a critical density factor $\epsilon=(1 + n_{\rm crit}/n)^{-1}$ depending on the critical density $n_{\rm crit}$, i.e. the ratio between the Einstein coefficient and the collisional de-excitation rate~\citep{Hollenbach1979}.  This critical density factor keeps into account the amount of energy which is radiated away and do not contribute to the heating.  We can define two extreme cases: if ({\em i}\/)  $n \gg n_{\rm crit}$,
then $\epsilon\rightarrow 1$, the gas is heated up by $\Delta H$; ({\em ii}\/) if $n \ll n_{\rm crit}$, then $\epsilon\rightarrow 0$,  and the energy $\Delta H$ is fully lost by radiative emission. The cases in between contributes to the heating of the gas to the extent of an amount of energy smaller than $\Delta H$.

\subsubsection{Photoheating}

The photodissociation and photoionisation induced by UV radiation generate an excess of energy equal to $h\nu-E_0$ which can go into heating, where $h\nu$ is the energy of the incident photon and $E_0$ the dissociation/ionisation energy threshold for the given species. Photoheating is mainly caused by ({\em i}\/) photoionisation of atoms in HII regions ($h\nu > 13.6$ eV), ({\em ii}\/) photoionisation of large molecules and small dust grains in HI regions ($h\nu <
13.6$~eV), and ({\em iii}\/) photoionisation of molecules in molecular regions. In a primordial environment the main heating source comes from photodissociation of H$_2$ and H$_2^+$, and under the presence of ionising radiation, from the photoionisation of H, and He. The heating is usually expressed as $\Gamma_\mathrm{ph} = H_\mathrm{ph} n_i$, where
\begin{equation}
H_\mathrm{ph} = \frac{4\pi}{h}\int_{E_0}^{\infty}\frac{J(E)\sigma(E)}{E}(E-E_0)\eta(E)\, dE
\end{equation}
is the photoheating rate in erg s$^{-1}$ for the given species. Its value depends on the radiation flux per energy $J(E)/E$, the cross-section $\sigma(E)$, the energy excess $E - E_0$, and the efficiency factor $\eta(E)$ that determines the amount of energy released into the gas.

\begin{figure}
\centerline{\includegraphics[scale=1]{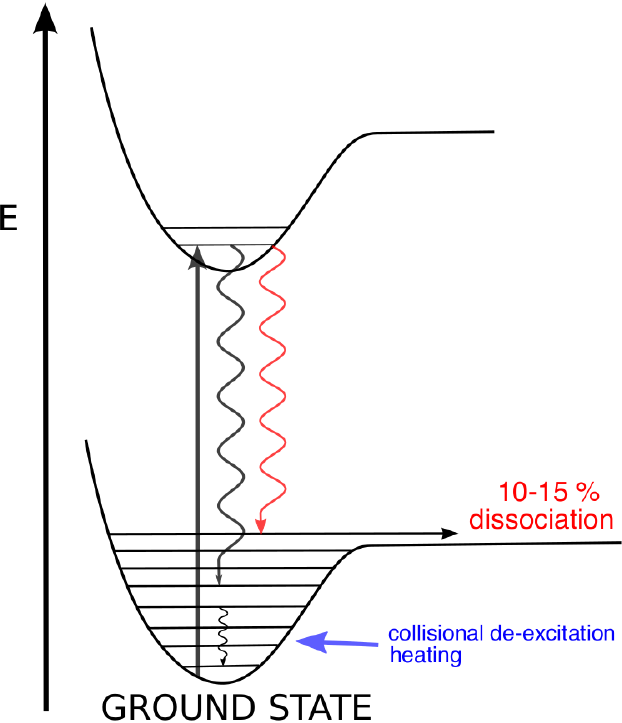}}
\caption{H$_2$ UV pumping heating sketch. Dissociation produces direct heating corresponding to 0.4~eV, while collisional de-excitations release up to 2.2~eV with an efficiency proportional to the critical density factor $\epsilon$ introduced in Sect.~\ref{sec:heatchem}.}
\label{fig:solomon}
\end{figure}

The heating produced by the photodissociation of H$_2$ discussed in Sec~\ref{sec:h2chem} is called {\em UV pumping heating} and it is the sum of two different contributions (see sketch in Fig. \ref{fig:solomon}). The first  (releasing~0.4 eV) is the heating produced by the dissociation of H$_2$ which directly goes into kinetic energy and has an efficiency $\sim 10$\%--15\%, as only this fraction of molecules dissociates following this path. The
second contribution (which releases~2.2~eV) is due to the de-excitation processes which involve vibrational excited states of the H$_2$ electronic ground state, and  depend on the critical density already introduced in sec.~\ref{sec:heatchem}. Photoheating can be the dominant contribution to the thermodynamics at low-densities where shielding effect are negligible and radiation can penetrate efficiently.

\section{Chemical uncertainties and dynamical implications}

Although the chemistry of the early Universe is bound to few elements (H, He, and D), and simple quantum molecular systems (H$_2$, HD), some of the ingredients leading the gas kinetics are affected by uncertainties. This in turn can alter the outcome of computational simulations of the formation and evolution of first stars and primordial black holes.

Particularly relevant are the uncertainties in studying the dynamical evolution of primordial halos irradiated by UV photons. The fate of the gravitational collapse is tied to the formation/destruction of H$_2$ and its capability to cool the gas down to $\sim 200$~K. The devoid of H$_2$ can strongly modify the outcome of the gravitational process and lead for example to the formation of super-massive black holes \citep{Omukai2001,Shang2010,Latif2014}. As discussed in this chapter the presence of a strong extragalactic UV background suppresses H$_2$ directly via photodissociation or indirectly through the photodetachment of H$^-$.  The minimum LW radiation flux required
to efficiently destroy H$_2$ and suppress its cooling is denoted $J_\mathrm{crit}$. Besides the chemical uncertainties on some of the rate coefficients, it has been shown \citep{WolcottGreen2011,Sugimura2014,Visbal2014,Agarwal2015} that the spectrum shape, and the treatment of the self-shielding are very  important for determining the critical flux. Several works reported values of $J_\mathrm{crit}$ in the range 20--$10^5$ (see \citet{Omukai2001,Shang2010,Inayoshi2011,Latif2014,Latif2015,Sugimura2014}) and large discrepancies have been found between simplified zero-dimensional model and more realistic three-dimensional hydrodynamical simulations \citep{Latif2014}. These differences span orders of magnitudes which are much more relevant than the factor of 2 uncertainties within the chemical model itself, as outlined e.g. in \citet{Glover2015a,Glover2015b}.

{
\acknowledgments
This work was (partially) funded by the CONICYT PIA ACT172033.
}

{
\bibliographystyle{ws-rv-har}    
\bibliography{ref}
}

\printindex[aindx]           
\printindex                  

\end{document}